# A comparison of the effects of different methodologies on the statistics learning profiles of prospective primary education teachers from a gender perspective

Jon Anasagasti* , Ainhoa Berciano , Ane Izagirre

Department of Didactics of Mathematics, Experimental Science and Social Sciences, University of the Basque Country (Universidad del País Vasco/Euskal Herriko Unibertsitatea), Basque Country, Spain
*Correspondence: jon.anasagasti@ehu.eus

## Abstract

Over the last decades, it has been shown that teaching and learning statistics is complex, regardless of the teaching methodology. This research presents the different learning profiles identified in a group of future Primary Education (PE) teachers during the study of the Statistics block depending on the methodology used and gender, where the sample consists of 132 students in the third year of the PE undergraduate degree in the University of the Basque Country (Universidad del País Vasco/Euskal Herriko Unibertsitatea, UPV/EHU). To determine the profiles, a cluster analysis technique has been used, where the main variables to determine them are, on the one hand, their statistical competence development and, on the other hand, the evolution of their attitude towards statistics. In order to better understand the nature of the profiles obtained, the type of teaching methodology used to work on the Statistics block has been taken into account. This comparison is based on the fact that the sample is divided into two groups: one has worked with a Project Based Learning (PBL) methodology, while the other has worked with a methodology in which theoretical explanations and typically decontextualized exercises predominate. Among the results obtained, three differentiated profiles are observed, highlighting the proportion of students with an advantageous profile in the group where PBL is included. With regard to gender, the results show that women's attitudes toward statistics evolved more positively than men's after the sessions devoted to statistics in the PBL group.

**Keywords**: Attitude, Learning Profile, Primary Education Undergraduate Degree, Project-Based Learning, Statistical Competence

As argued in the Guidelines for Assessment and Instruction in Statistics Education II (GAISE II) (Bargagliotti et al., 2020), the demands for statistical literacy have never been greater. For example, students who finish the secondary school should develop the ability to interpret, critically evaluate, and communicate about statistical information (Gal, 2002), with the ultimate goal of understanding decision making under uncertainty (Shaughnessy, 2019). To achieve this objective, the educational work should begin in Childhood Education and should continue throughout Primary Education, taking into account certain key ideas for improving teaching (Alsina et al., 2023): involving students in the development of projects in which they have to collect their own data, organising the data collected in frequency tables, encouraging the use of different graphs, posing critical questions to students that invite them to maintain a reflective stance or encouraging the learning of techniques for numerically interpreting the data represented. All these ideas can be developed through meaningful projects or investigations in the manner, for example, of the PPDAC (Problem, Plan, Data, Analysis, and Communication) model proposed by Wild and Pfannkuch (1999).

In the same way, future primary school teachers must have learning programmes that are themselves adequate examples of good didactics (Garfield & Everson, 2009). In this sense, we start from a concrete Project Based Learning (PBL) approach proposed by Anasagasti (2019), which follows the recommendations laid down in the GAISE (Franklin et al., 2007); that is, emphasising statistical knowledge and developing statistical thinking, using real data, reinforcing conceptual understanding beyond mere procedural learning, encouraging active learning in the classroom, using technology for the development of conceptual understanding and data analysis, and using assessment methods for the improvement and evaluation of student learning.

There are many voices recommending the use of projects to work on statistics; as Moore (2005) mentions, there is evidence that "active learning" strategies are superior to the "information transfer" model that underlies much traditional instruction. Apart from being able to improve the different techniques of statistical calculation or graphical representation, as occurs in classes where classroom dynamics are focused on theoretical explanations and exercises, conceptual reasoning and data interpretation are favoured to a greater extent through contextualised projects (Alsina et al., 2023). In this sense, we found different works with future teachers of both childhood (Berciano et al., 2021), primary (Rivas et al., 2019; Ubilla & Gorgorió, 2020, Anasagasti et al., 2022) and secondary school (Makar and Confrey, 2004; Garfield & Everson, 2009), where projects and statistical research problems, in addition to being good examples from a didactic point of view, play an important role in developing the standards defined by the National Council of Teachers of Mathematics (NCTM) for the data analysis and probability block (Metz, 2010).

In addition to the recommendations for the use of projects from a competence point of view, it is also interesting to note the suggestions made from an attitudinal point of view. The relationship between the acquisition of certain competences and the attitude a person has towards tasks related to these competences seems undeniable from today's perspective. Deficits in functioning in situations related to mathematics in general and, more specifically, to statistics, can create a vicious circle in which the avoidance of challenging stimuli related to such knowledge will create gaps in learning. In this regard, studies such as that of Ashaari et al. (2011) indicate that students feel quite intimidated, fearful and stressed when solving problems related to statistics.

In the case of future teachers, this is crucial, since if a teacher is insecure about his or her statistical knowledge, it may lead to students progressing with an inadequate perception of their ability to learn (Sánchez-Mendías et al., 2020). Specifically, in the field of statistics and in relation to future primary school teachers, we find studies that, while focusing on competence (sometimes understood as knowledge) and attitudes towards statistics, relate both concepts (Estrada et al., 2017; Anasagasti, 2019). The improvement of these attitudes often requires the improvement of competence; as Estrada et al. (2017) indicate, "The shown effect of knowledge on attitudes suggests that better preparation of teachers is a prerequisite if we want to improve their attitudes" (p.87).

Although these relationships between competences and attitudes are of great interest as they can offer us the possibility of orienting the teaching of statistics in one direction or the other, as far as the authors know, they have not been studied; so the categorization of different learning profiles involving both is of fundamental relevance in order to provide results that reveal how to approach the teaching-learning of statistics from a more holistic point of view.  Furthermore, when referring to learning profiles related to the study of mathematics or statistics, we find in the literature different approaches depending on, for example, teaching-learning styles (Santoyo et al., 2017), emotions and performance (Sachisthal et al., 2021), cognitive engagement and study burnout (Asikainen et al., 2020), professional competences (Muñiz-Rodríguez, 2020), or the participants' attitude towards the subject (Zamalia, 2009). So, in this study we proceed to identify different profiles according to the two variables mentioned: statistical competence and attitude towards statistics.

It is of great importance to focus on how to effectively introduce statistics in the Primary Education (PE) classroom and, therefore, as a didactic model, in the university classroom of future teachers. As suggested by Estrada et al. (2011), students learn statistics more effectively in environments where collaboration is encouraged and through progressive teaching methods such as discovery learning and problem solving. For this reason, when identifying the different profiles, we want to introduce the methodology used with future PE teachers as a study variable. As has been indicated, there are more and more studies that recommend introducing these contents through active methodologies that make it possible to create rich learning contexts for students (Batanero & Díaz, 2004); specifically, it is of great interest to set up statistical research projects with future primary school teachers (Ubilla, 2019). In this specific case, the aim is to study whether the use of Project Based Learning (PBL) methodology, in addition to serving as an example for the practice of future teachers, is related to the students adopting one type of profile or another.

Finally, we want to analyse the results from a gender perspective and see if there is a difference according to the gender of the participants, since there are studies that claim that women's attitude towards mathematics in general is worse than men's (Daches Cohen et al., 2021) and that, regardless of grades, they have a higher risk of school burnout (Herrmann et al., 2019). This may be related to a lower rate of women pursuing STEM (Science, Technology, Engineering and Mathematics) studies.

The objectives of this research are, therefore, on the one hand, to obtain and describe statistics learning profiles in terms of statistical competence development and the evolution of attitudes towards statistics, highlighting the characteristics of the corresponding exit profiles. On the other hand, this research aims to study whether there are differences in the profiles obtained depending on the methodology implemented, comparing a group that works through Project Based Learning (PBL) with another group where theoretical explanations and typically decontextualized exercises predominate. Finally, the aim is also to analyse whether the gender of the participants can be a variable related to whether they belong to one learning profile or another.

## METHODS

### Research Design

This study, which is exclusively quantitative in nature, is based on a cluster analysis for which the variables "statistical competence development" and "evolution of their attitude towards statistics" are taken into account; the questionnaires used to measure these variables are detailed in the Research Instrument section below. Once each profile has been defined, it returns which students belong to each of the defined profiles and, in this way, the mean values of the variables mentioned are obtained, as well as other characteristics of the persons belonging to each profile (e.g., group to which they belong or gender).

Bearing in mind that the study also presents an analysis of the relationship between the different profiles and the methodology implemented in the classroom, it is a quasi-experimental design in which the groups were not randomly assigned but were established beforehand. However, the division of the two groups is for organisational reasons and is based on the alphabetical order of the surname. For this reason, it is understood that, despite not having a strictly experimental design, the research and control groups correspond to the characteristics of a balanced design. This fact is confirmed by the fact that no statistically significant differences were found between the two groups (pre-test) in statistical competence or in attitude towards statistics (Anasagasti, 2019).

## Participants

The sample was selected for convenience within the Faculty of Education of Bilbao at the University of the Basque Country (UPV/EHU) because of the availability and willingness of the participants to take part in the study; they were assured that the processing of their data would have no impact on their academic results and that the processing of their data would be anonymous. Bearing in mind that the presence of the students in all the sessions in the Statistics block was essential for data collection, the final sample consisted of 132 participants, 69 in the research group and 63 in the control group. In total, there were 50 men and 82 women, with an average age of 21 years (with a standard deviation of 3.61 years). The gender distributions within each group are presented in Table 1.

**Table 1**. Distribution of participants according to gender and group

|  | Men | | Women | | Total |
|---|---|---|---|---|---|
|  | Frequency | Percentage | Frequency | Percentage | Frequency |
| Research G. | 26 | 37.7% | 43 | 62.3% | 69 |
| Control G. | 24 | 38.1% | 39 | 61.9% | 63 |
| Total | 50 | 37.9% | 82 | 62.1% | 132 |

The period dedicated to the study of the statistics block was carried out during the teaching hours of the compulsory third-year subject "Mathematics and its Didactics II" of the Degree in Primary Education, which includes the block "Information processing, chance and probability". The duration of the interventions, both for the research group and the control group, was limited to five weeks, in which each student had a total of six hours of theory classes and seven and a half hours of practical classes.

As for the design of the intervention carried out with the research group (ABP), detailed in the work of Anasagasti and Berciano (2016), a distinction is made between the presentation activities, the traditional activities that are mainly developed during the theoretical classes, and the different activities that make up the implementation of a project structured according to the recommendations of GAISE (Metz, 2010) in which the students must solve a research problem. In the case of control group, the methodology carried out has been the usual teaching-learning process of statistics of the last several years at the faculty, based on theoretical lectures explaining concepts, examples and classical exercises (not contextualized) related with them. To ensure that the theoretical design and implementations were aligned, the research group coordinated with the teachers of both groups at all times.

## Research Instrument

To measure each of the two variables to be taken into account in the cluster analysis, two questionnaires were used; the students answered these both before and after the period in which they worked on the subject. The Test of Statistical Competence (Anasagasti, 2019) was used to measure statistical competence, and the Survey of Attitudes Towards Statistics (Schau, 2003) was used to measure attitudes towards statistics.

The first of these, the Test of Statistical Competence (Anasagasti, 2019), is a questionnaire designed specifically for the framework research and takes a large part of its items from previously validated questionnaires as the Statistics Reasoning Assessment (Garfield, 2003). Heterogeneous in nature as the items cover different concepts, from 17 questions composed of 50 items measuring different aspects of statistical competence (minimum score: 0; maximum score: 50), the following values are obtained for Cronbach's Alpha coefficient: α=.573 in the pre-test and α=.579 in the post-test. This value is moderate and it is considered enough in this case, because of the nature it has, based on the definition of its sub-competences. Four sub-competences are analysed:

- Statistical knowledge: knowing, analysing and applying basic knowledge for classroom practice in PE, both in its conceptual and didactic aspects.
- Knowledge of the Primary Education Curriculum: Knowledge of the school mathematics curriculum and its application to the analysis of proposals for the area in PE and to the design and development of new activities.
- Technological knowledge: Knowledge and appropriate use of teaching materials and technological media to model different learning situations.
- Knowledge of the usefulness of statistics: Recognise the role of mathematics as a fundamental element in the development of logical thinking, precision, rigour and the ability to evaluate decisions.

The second test, the Survey of Attitudes Towards Statistics (Schau et al., 1995), assesses the degree of positive attitude of each person towards statistics. This test has been validated in previous research and, in this case, has been translated into Basque and validated by bilingual experts in the field of mathematics teaching. The test consists of 28 Likert-type items that are grouped into four components so that, once all the items have been answered, the scores obtained in each component are known. To assess the overall attitude of each person towards statistics, as presented in studies such as Estrada et al. (2017), the scores of the four components are added to create a new variable that measures the total score (minimum score: 28; maximum score: 140). This instrument has been validated in several previous works, and obtained in the present sample a Cronbach's Alpha coefficient for the total scale of α=.880 in the pre-test and α=.889 in the post-test. The components are defined as:

- Affective Component: positive or negative feelings towards statistics.
- Cognitive Competence Component: Perception of one's own capabilities in statistical knowledge and skills.
- Value Component: Usefulness, relevance and value attached to statistics in personal and professional life.
- Difficulty Component: Perception of ease or difficulty of the subject of statistics.

## Data Collection and Analysis Procedure

In order to obtain the different learning profiles during the study of the statistics block, the results obtained in the questionnaires measuring the statistical competence and attitude towards statistics of future primary school teachers were used as a starting point. In order to obtain the development or evolution experienced during the study period, we have defined this as the difference in score (post-test minus pre-test) in each questionnaire. Once these results were obtained, a cluster analysis was carried out according to the two variables mentioned (statistical competence development and evolution of their attitude towards statistics) using the K-means algorithm, forming a single partition without hierarchizing or relating the clusters to each other (Pérez, 2004). These analyses were carried out using SPSS Statistics, version 24.

In order to compare these profiles according to the teaching methodology implemented during the period of study of the "Information processing, chance and probability" block, two groups of students have been differentiated: the research group (with PBL methodology) and the control group (with methodology based on theoretical explanations and exercises). The other variable to be considered in this research was also recorded by means of questionnaires: the gender of the participants.

## RESULTS AND DISCUSSION

The results corresponding to the aspects detailed in the research objectives are shown below. Firstly, the way in which the different learning profiles were obtained is explained; then, each of them is described in terms of the two main variables and their corresponding output profiles; finally, the distribution of the people participating in the study is shown in terms of the other variables considered: methodology group and gender.

### Obtaining the Learning Profiles

In order to facilitate the interpretation of the results, and bearing in mind that the aim is to determine principal characteristics for a number of groups that will allow the future implementation of a teaching-learning process adapted to each case, but also methodologically feasible, it has been decided to limit the number of theoretical groups to a maximum of 5. After performing several clustering tests with 3, 4 and 5 clusters and with the intention that they should be formed by people with similar characteristics, small variance between them and a similar number of people in each cluster, it was finally decided to classify the students into 3 clusters. In this way, the groups or clusters obtained are intended to be as homogeneous as possible, maximising the distance between their centres of gravity, i.e. the variance between the different clusters, and minimising the variance of the variables observed within each one. The exact scores for each cluster and the number of people assigned to each cluster are shown in Table 2.

Table 2. Cluster composition and characteristics

| Cluster | N | Evolution on TSC | | Evolution on SATS | |
|---------|-----|-------|-------|--------|--------|
|         |     | Mean  | S.D.  | Mean   | S.D.   |
| 1       | 50  | -2.32 | 3.178 | 4.68   | 5.516  |
| 2       | 43  | 5.42  | 2.93  | 8.32   | 7.846  |
| 3       | 39  | 1.49  | 2.981 | -12.49 | 10.27  |
| Total   | 132 | 1.33  | 4.437 | .80    | 11.773 |

In order to be able to observe more clearly the differences between the obtained profiles, Figure 1 shows a correspondence between the evolution experienced by each cluster in both statistical competence and attitude in their absolute values shown in the third and fifth columns where the means of these changes are described, as well as the size of each cluster, written in the second column (N).

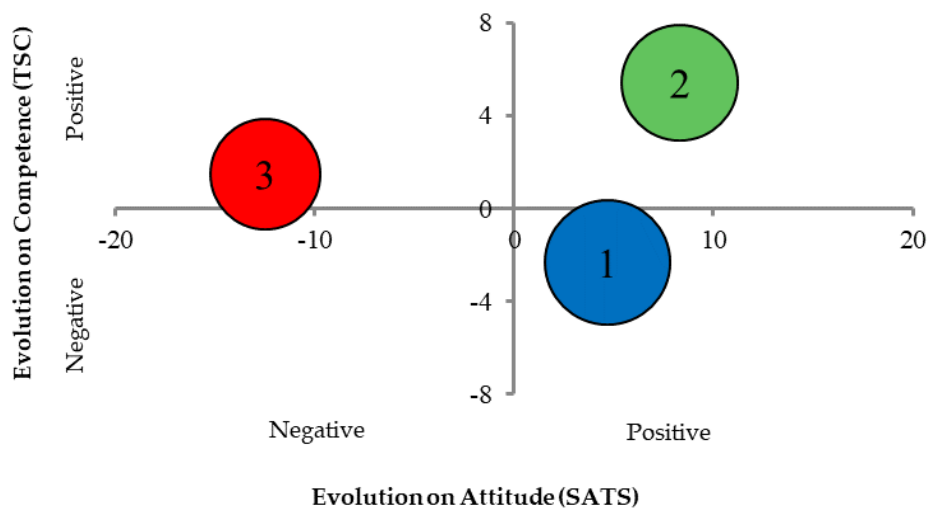

**Figure 1.** Correspondence between evolution in Statistical Competence (TSC) and in Attitude (SATS).

As described above, the range of TSC scores is 50 points (between 0 and 50); in the case of SATS, the range is 112 (between 28 and 140). Looking at the mean of the whole student data set, when comparing the evolution of the TSC and SATS tests, we observe improvements in in both cases (1.33 for TSC test, 0.8 for SATS test), although this improvement is moderate. However, once the cluster scores are obtained, the positive and negative trends become more evident, with much more noticeable inter-group differences being observed in the case of attitudes at both absolute and relative levels.

Broadly speaking, these clusters enable us to describe the profiles of the students as follows:

- First Profile (1): determined by Cluster 1, students who do not evolve positively or evolve slightly negatively in terms of statistical competence (TSC), but achieve a slightly positive evolution in terms of attitude (SATS).
- Second Profile (2): determined by Cluster 2, students with a significant positive evolution in both questionnaires.
- Third Profile (3): determined by Cluster 3, students who achieve a slight positive evolution in their statistical competence, but evolve considerably in a negative direction in attitude (-12.49 with respect to a range of 112).

Statistical competence and attitude towards statistics appear to be related, as shown by previous studies such as that of Estrada et al. (2011), who point out that negative attitudes seem to be linked to the perceived difficulty, lack of knowledge and excessively formal content of the relationship between the two. However, in this study, the profiles have been defined on the basis of the evolution experienced in these variables after the statistics course and, consequently, regardless of whether they present greater or lesser competence or a more or less positive attitude towards statistics, what has been measured is the change experienced after the course.

## Description of Profiles and Output Profiles

Once the different learning profiles have been obtained, we are interested in knowing their main characteristics, paying special attention to their output profiles, detailing the final scores (post-test) obtained in each sub-competence of the TSC and in each component of the SATS. In order to show more clearly the main differences between sub-competences and components of each cluster, the scores are re-scaled from 0 to 10 (Figure 2). This allows us to observe more easily the differences between the sub-competences and components of each cluster, and to observe those areas where prospective teachers seem to present greater competence difficulties or more negative attitudes towards statistics.

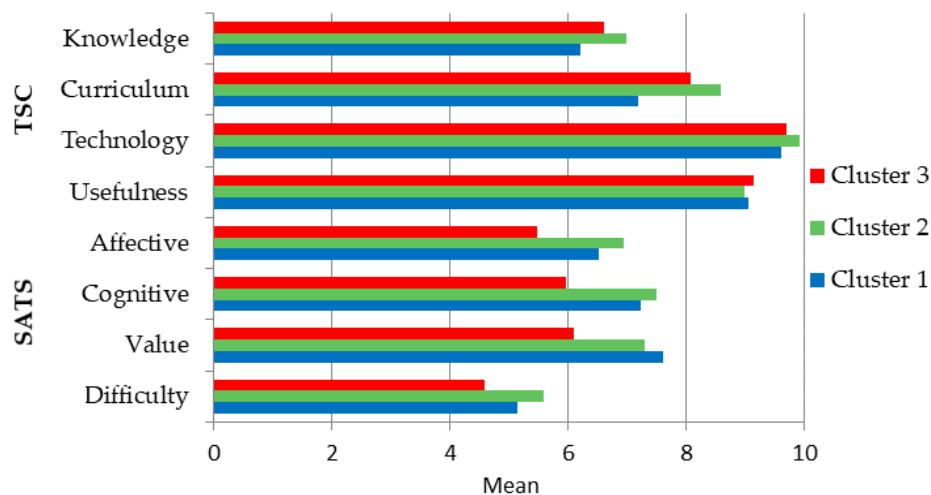

**Figure 2.** Normalized means obtained in each sub-competence (TSC) and component (SATS)

In Profile 1 (determined by cluster 1), we find students who barely manage to improve or even partially worsen their statistical competence. As can be seen in Figure 2, while in the sub-competences of Technological Knowledge and Utility they achieve similar scores to the other profiles, the sub-competences of Statistical Knowledge and Primary School Curriculum Knowledge obtain considerably lower scores; this could be interpreted as meaning that they mainly fail to respond to concrete concepts or memorisation. In terms of attitude, they improve it (+4.68, see Table 2), that is, they obtain better results than the average of the evolution of the attitude of the total group (+.80), standing out from the other clusters in the Value component. In summary, these are students who present a profile with a fairly good attitude towards statistics, understand the value and usefulness it may have, but do not advance in their competence mainly due to a lack of conceptual understanding.

One of the objectives of the instruction carried out is to improve statistical competence and this is the only profile that does not achieve this, so it seems obvious that we should focus especially on the conceptual improvement of these future teachers. Moreover, although they have a good attitude towards the subject, it should not be overlooked, as studies such as Asikainen et al. (2020) indicate, that students who have problems in mastering and understanding the core knowledge, as well as having fragmented knowledge about the subject end up feeling more exhausted, less capable and more cynical about their learning in the long run.

Profile 2 contains those who make progress in competence and attitude and stand out positively from the rest of the profiles in all sub-competences except Knowledge of Usefulness, and in all components except Value (Figure 2). This Profile 2 is therefore of great interest to us, since the desired effect or result is achieved on them, a balanced improvement in competence and attitude towards statistics. About the teacher's response in future interventions, in terms of the exit profile, it would be valuable to continue working on the usefulness and value of statistics through new research proposals or even other types of projects.

Finally, Profile 3 contains students who, despite achieving a small improvement in statistical competence, register a considerable worsening of their attitude. This group is situated between the other two profiles in three of the defined sub-competences and even surpasses them in Utility Knowledge, but in the SATS, it is evident that the final scores obtained are considerably lower than those of the other two clusters, achieving a score of 6 in only one component. In short, this is a profile of students who, despite managing to improve their competence and to understand the usefulness of statistics, have an attitude that worsens considerably. This may be because they do not value this usefulness sufficiently, because they do not like the subject or because they think that it is a complicated subject that they are not good at.

For this reason, it seems advisable to consider making various methodological changes. As indicated in the study by Zamalia (2009), where students are differentiated according to whether they have a positive or negative attitude towards statistics, although, in general, the proportion of students with a negative attitude is small, instructors should be aware of the effect of their methods and approach to teaching statistics on students' attitudes towards statistics.

### Profiles according to the Group to which They Belong (Research/ Control)

First of all, it should be noted that prior to the implementations carried out, there are no statistically significant differences between the research group and the control group in either the TSC or the SATS, thus ensuring that we start from a comparable situation (Anasagasti, 2019). Analysing the results with respect to the two methodology groups (i.e. those who have worked either with PBL, or alternatively, with theoretical explanations and decontextualized exercises), there is an unequal proportion between the number of students in each profile. This can be seen in Figure 3.

Of the research group, 38% are classified in Cluster 1, 46% in Cluster 2, and 16% in Cluster 3. On the other hand, in the control group, 38% are classified in cluster 1, 18% in cluster 2, and 44% in cluster 3. It can be seen that, although in Profile 1 the proportion of students is practically the same according to the group to which they belong, in the second and third profiles the difference in proportions is remarkable.

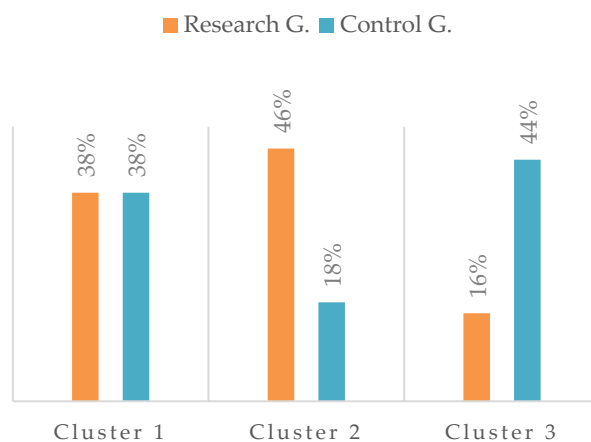

**Figure 3.** Percentage of students of each cluster according to methodology (Research/Control).

Within the research group, there is a higher proportion of students belonging to Profile 2 who show a positive evolution in the two questionnaires, while in the control group there is a higher proportion of students classified in Profile 3, in which they evolve negatively in terms of attitude and improve little in statistical competence. Given this difference, it seems clear that the PBL group has a higher proportion of students who improve their attitude towards statistics, obtaining similar results for statistical competence.

The differences in the number of students who are classified within methodology group leads us to conclude that PBL methodology promotes to a certain extent a greater evolution of the students in Profile 2 than in Profile 3. We understand that in this way learning is achieved which, while generating a benefit in competence similar to that achieved through traditional methodology, produces a considerably greater benefit in attitude, thus reducing study-related burnout and promoting the well-being of the students. We agree with Estrada et al. (2011) that one of the main influences on teacher attitude, in addition to prior knowledge of statistics, is good prior learning experiences. Therefore, as the authors indicate, it is the responsibility of university teachers to create an emotionally and cognitively supportive environment where future primary school teachers gain confidence in their own ability both to learn and to learn to value the role of statistics.

One of the factors that we consider it is important to take into account when interpreting that PBL methodology achieves a greater number of people who are classified within Profile 2, is the fact that students get involved in developing research cycles, specifying their own initial research question or using real data obtained by themselves, thus helping, as Ubilla and Gorgorió (2020) point out, to give meaning to the study of statistics. Furthermore, we assert, as Perrenet et al. (2000) indicate, that this type of methodology promotes the autonomy and self-direction that students take. This is important if students are to continue developing new research proposals. In undergraduate studies, future teachers will have at least one opportunity to implement research proposals in their final projects, so we believe that this task can also be of great benefit to their academic development.

## Gender Profiles

Finally, the allocation of students to the different clusters according to gender is analysed below. Taking into account the 132 people in the study, Figure 4 shows that the percentage of men is evenly distributed among the three clusters. For women, however, the distribution differs by cluster, with a higher percentage of women in Cluster 1 (41% women versus 32% men), a similar percentage of men in Cluster 2 (both around 33%), and a lower percentage in Cluster 3 (27% women versus 34% men). These data suggest that, irrespective of the methodology used in class, the proportion of women improving their attitude towards statistics is higher than that of men, although their competence does not improve considerably.

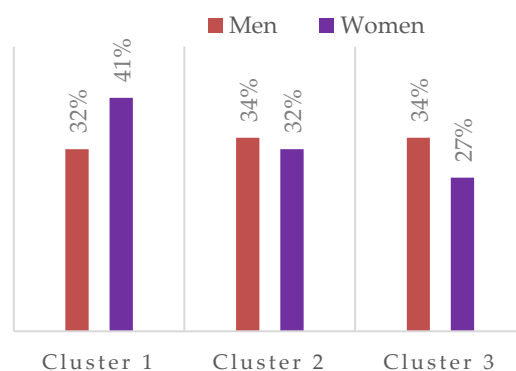

**Figure 4.** Percentage belonging to each cluster according to gender (132 students).

Looking only at the participants of the research group, the proportions are distributed differently from those of the students as a whole (Figure 5). In the case of men, a large percentage (half of all men) are in Cluster 2, with the vast majority achieving favourable progress in both statistical competence and attitude. Among women, there is an equal distribution between Clusters 1 and 2 (44% in each), in which the common characteristic is that they evolve positively in attitude.

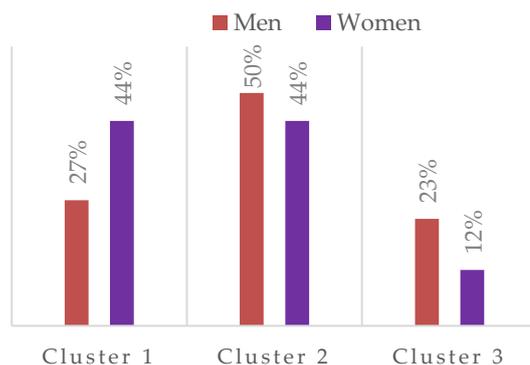

**Figure 5.** Percentage belonging to each cluster according to gender (research group).

For both genders, the lowest proportion of students, 23% of men and less than 12% of women, belong to Cluster 3, where they perform positively on competence, but negatively on attitude. Taking into account recent studies such as Daches Cohen et al. (2021), which indicate that information related to mathematics is linked to a more negative affectivity in women, in this study we also wanted to observe the difference in the proportions of belonging to each profile according to the gender of the participants. As a clear consequence of this distribution of profiles, it seems that women's attitudes towards statistics evolve more positively than men's after the sessions devoted to it. This coincides with studies such as that of Fullerton and Umphrey (2001) where it is pointed out that, after doing a course in statistics, women are generally more positive in most of those components pertaining to attitude towards the subject. This could be understood from the point of view that women turn their anxiety into something positive by being more inclined to seek help from their instructor and ask questions in class. This happens even more so in the case of introducing active methodologies that favour meaningful and contextualised learning, as indicated by the data (only 12% of the women in the research group are in Profile 3).

## CONCLUSION

Diversity is a very important factor to take into account if we want to improve competence and attitude towards any subject. The characterisation of the profiles of future teachers can help us to understand in a general way the needs that each student profile may have. In this sense, we would mention that this is the first study that has been done to characterize these profiles taking into account both, competence and attitude. Therefore, the results of this study allow us to respond to the objective of this work, to analyse the learning profiles associated with the teaching methodology of statistics, giving us the opportunity, on the one hand, to evaluate in what sense the achievements of university work influence future primary school teachers and, on the other hand, to know the needs that different student profiles may have during and after a specific intervention in statistics. We have proved that there are three clear learning profiles, where the most suitable, the one that improves competence and attitude towards statistics, is mainly when the methodology used is PBL. Likewise, the one who obtains the worst results is the one with a more masterly methodology. With respect to the gender perspective, it has been seen that the teaching methodology used has different repercussions on statistical competence and attitude towards statistics, i.e., in case where PBL is included, the proportion of women with an advantageous profile is considerably higher than in the other methodology. In fact, they are generally more positive in most of those components pertaining to attitude towards the subject in PBL case.

Anyway, as limitations of this study, we should mention that one of them is that the data collection about attitudes is based on participants' perceptions, i.e., indirect measures. Another one is that it has not been possible to analyse the reasons for these perceptions. As future research lines, we have that, based on the identification of profiles and the analyses presented in this study, it would be interesting to determine and evaluate the implications of the specific implementations designed according to the needs of each profile. It would be appropriate to contrast the results about attitudes with another instrument of analysis, for example, based on the observation of their own teaching practice during the training period. In this line, it is necessary to go deeper, using qualitative group methodologies, into the causes that lead to a relatively low appraisal of their development of certain competencies by the people classified within the third profile. Finally, we must conclude by highlighting the results of this article, because, as has been said, they have allowed us to make progress in the knowledge of the factors that determine the learning of statistics to be more meaningful.

## Acknowledgements

Article partially funded by the KOMATZI Research Group (GIU21/031) of the University of the Basque Country (Universidad del País Vasco/Euskal Herriko Unibertsitatea).

## Declarations

| Author Contribution | : | JA: Conceptualization, Methodology, Investigation, Writing - Original Draft, Editing and Visualization. |
|---|---|---|
| | | AB: Methodology, Formal analysis, Writing - Review & Editing, Funding acquisition, and Supervision. |
| | | AI: Formal analysis, Writing - Review & Editing, Visualization. |
| Funding Statement | : | This research was partially funded by the KOMATZI Research Group (GIU21/031) of the University of the Basque Country (Universidad del País Vasco/Euskal Herriko Unibertsitatea). |
| Conflict of Interest | : | The authors declare no conflict of interest. |
| Additional Information | : | Additional information is available for this paper. |

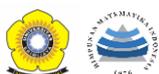

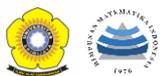